\documentclass[aps,twocolumn,prl,showpacs]{revtex4}
\usepackage{graphicx}

\newcommand{\be}{\begin{equation}}
\newcommand{\ee}{\end{equation}}
\newcommand{\bea}{\begin{eqnarray}}
\newcommand{\eea}{\end{eqnarray}}

\begin{document}

\title{Giant light shift of atoms near optical microstructures}

\author{Peter Horak}
\author{Peter Domokos}
\altaffiliation{On leave from: Research Institute for Solid State Physics and
Optics, Hungarian Academy of Sciences, Hungary}
\author{Helmut Ritsch}

\affiliation{Institut f\"ur Theoretische Physik, Universit\"at Innsbruck,
Technikerstra{\ss}e 25, A-6020 Innsbruck, Austria}

%\date{\today}

\begin{abstract}
\vspace{3mm}
Atoms coupled to optical fields confined in one and two spatial dimensions in
solid state microstructures can experience very large light shifts if the
driving frequencies are close to a resonance of the microstructures and an
atomic transition. Using the simple example of a quasi one-dimensional
waveguide structure we can analytically calculate the atomic AC Stark shift and
the modifications of the light field induced by the presence of the atom. A
large enhancement of the effective interaction strength is found due to a non
uniform mode density.
Experimentally this should be visible by monitoring the scattered light field
as well as by the modification of the atomic trajectories bouncing from the
evanescent light.
\end{abstract}

\pacs{03.75.Be, 42.25.Bs, 42.82.Et}

\maketitle

In the past years we have seen spectacular advances in our ability to
cool atoms to nanokelvin temperatures and control their motional
degrees of freedom down to the quantum level \cite{BEC}. In parallel,
the miniaturization of optical microstructures has reached the level
where fabrication almost at the atomic scale is feasible
\cite{nanooptics}. One of the great goals in the near future is to
bring together these technologies in a generation of integrated
optical quantum devices \cite{Joerg}. As a central point to utilize
such devices, we must understand the behaviour and transport of the
atomic matter subject to subwavelength-structured electromagnetic
radiation fields. Evanescent fields created by tailored dielectric
microstructures are of primary interest in this research. A series of
experiments have demonstrated the usefulness of evanescent optical
fields to realize atom mirrors \cite{Aspect} or quasi 2D surface traps
\cite{Grimm}. However, these surface setups have all been based on
macroscopic fields regarding the transverse spatial dimension and
photon numbers involved. Hence a single particle has almost no effect
on the field and a single photon field would give only a negligible
force on an atom. In this Letter we show that the effective
atom-photon interaction can strongly be enhanced for atoms at the
surface of dielectric microostructures, where the field is partly
confined in some directions. Note that for a full 3D confinement one
would recover a setup as in cavity QED \cite{Brune} with special
boundaries. Here we reveal new surprising phenomena beyond the 
effects of a single strongly coupled radiation mode. For example, a
continuum of travelling wave modes can induce a huge atomic AC Stark
shift exceeding by orders of magnitude the natural linewidth. In
conjunction with the large light shift, a strongly increased photon
scattering rate by a single atom takes place, which could be used for
position and state selective single atom detection and manipulation
schemes \cite{Domokos}.

%%%%%%%%%%%%%%%%%%%%%%%%%%%%%%%%%%%%%%%%%%%%%%%%%%%%%%%%

\textit{Atomic light shift in 1D continua of modes.-}
Let us consider a two-level atom with resonance frequency $\omega_a$ and
free-space spontaneous emission rate $\Gamma$ in (or close to) a dielectric
medium with refractive index $n_0$ which is assumed to be infinitely extended
into the $z$ direction, but with a transverse dimension of the order of an
optical wavelength. This microstructure supports optical modes which are
described by annihilation operators $a_n(k)$, where $n$ labels the transverse
mode index and $k$ the longitudinal wave number. The annihilation and
creation operators fulfill the standard commutation relation
\be
\left[a_n(k),a^\dagger_{n'}(k')\right] = \delta_{nn'}\delta(k-k').
\ee
The corresponding frequencies are denoted by $\omega_n(k)$ and the mode
functions read  $f_n(k,\mathbf{x}) = \exp(ik z) f_n^{(T)}(k,x,y)$. 
The mode functions are normalized such that
\bea
A & = & n_0^2\int_{A_1}dx\, dy |f_n^{(T)}(k,x,y)|^2 \nonumber \\
 & & +\int_{A_2}dx\, dy |f_n^{(T)}(k,x,y)|^2,
\eea
where the first integral goes over the part of the mode function \textit{inside}
and the second integral over the part \textit{outside} the dielectric medium.
$A$ is the cross section of the microstructure. The positive frequency part of
the electric field is then given by
\be
E^{(+)}(\mathbf{x}) = \sum_n \int dk E_0(\omega_n(k)) f_n(k,\mathbf{x}) a_n(k)
\label{eq:field}
\ee
with $E_0(\omega) = \sqrt{\hbar \omega/(2\epsilon_0 A)}$ the electric field of
a single photon.

In the following we will assume that the medium is pumped by monochromatic
light of frequency $\omega_p\approx \omega_a$ and that only modes with 
frequencies
close to this contribute to the system dynamics. We will thus replace
$E_0(\omega_n(k))$ by the corresponding value at the pump frequency
$E_0 = E_0(\omega_p)$ and pull it out of the integral in
Eq.~(\ref{eq:field}). In dipole and rotating wave approximation and in a
frame rotating with $\omega_p$ the total system dynamics is then governed by the
Hamiltonian
\bea
H & = & -\Delta_a \sigma^\dagger\sigma 
        -\sum_n\int dk \Delta_n(k) a^\dagger_{n}(k)a_{n}(k) \nonumber \\
& & + i g\sum_n\int dk \left[f_n^*(k,\mathbf{x}_a)a^\dagger_{n}(k)\sigma
         -\sigma^\dagger f_n(k,\mathbf{x}_a)a_{n}(k)
     \right] \nonumber \\
& & +i \sum_n\int dk \left[\eta_n(k)a_{n}(k)-\eta^*_n(k)a^\dagger_{n}(k)
     \right].
\label{eq:ham}
\eea
Here $\sigma$ is the atomic lowering operator,
$\Delta_a=\omega_p-\omega_a$, $\Delta_n(k)=\omega_p-\omega_n(k)$,
$\mathbf{x}_a$ is the position of the atom, and $g=\mu E_0$ where
$\mu$ is the atomic dipole moment. The first line of
Eq.~(\ref{eq:ham}) describes the free field evolution, the second line
the atom-light coupling, and the third line the coherent pumping of
the light modes.

Allowing for photon losses from the dielectric medium at a rate $2\kappa$ and
assuming that the dynamics of the electric field occurs on a much faster time
scale than the atomic dynamics, we may adiabatically eliminate the photon
operators from the Hamiltonian. In the limit of small atomic saturation and
neglecting quantum noise terms, we finally obtain the Heisenberg equation of
motion for the atomic internal variable $\sigma$,
\be
\frac{d}{dt}\sigma = (i\Delta_a-\Gamma + L)\sigma - \chi
\ee
where
\bea
L & = & g^2\sum_n\int dk \frac{|f_n(k,\mathbf{x}_a)|^2}{i\Delta_n(k)-\kappa}, 
\label{eq:shift}\\
\chi & = & g\sum_n\int dk
\frac{f_n(k,\mathbf{x}_a)\eta_n^*(k)}{i\Delta_n(k)-\kappa}.  
\eea 
We therefore see that the coupling of the atom to the field modes
gives rise to a driving term $\chi$ of the atomic operator $\sigma$
and, additionally, to an atomic light shift $L$. In general, $L$ is
complex and the atom will thus experience a frequency shift according
to the imaginary part of $L$ and a resonance line broadening described
by its real part. Depending on the functional dependence of
$\Delta_n(k)$, the light shift obtained from the coupling of the atom
to continuous one-dimensional sets of electric modes can significantly
alter the atomic dynamics, especially if $\Delta_n(k)$ is a highly
asymmetric function. Since dielectric microstructures can be
fabricated with high accuracy, this offers the possibility to tailor
the atom-light interaction to a large extent.

%%%%%%%%%%%%%%%%%%%%%%%%%%%%%%%%%%%%%%%%%%%%%%%%%%%%%%%%

\textit{Specific example of a microstructure.- }
In order to discuss this effect quantitatively in more detail, we will
in the following concentrate on a specific example of such a
microstructure.

\begin{figure}[tb]
\includegraphics[width=15em]{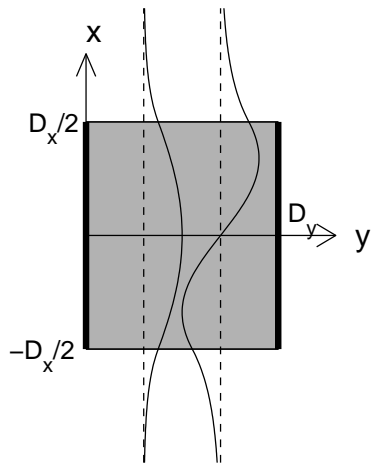}
\caption{Schematic presentation of the sample microstructure.}
\label{fig:structure}
\end{figure}

Let us consider a dielectric medium with a rectangular cross section
of height $D_x$ and width $D_y$ as sketched in
Fig.~\ref{fig:structure}, but infinitely extended in the $z$
direction. The height $D_x$ is chosen such that for large $D_y$ two
transverse modes are supported at the pump frequency $\omega_p$ as
indicated in the figure. (For simplicity we neglect polarization
issues here and therefore assume a scalar electric field.)  The
dielectric surfaces perpendicular to the $y$ axis, on the other hand,
are supposed to be coated with a highly reflecting metallic layer, and
the width $D_y$ is chosen small enough such that only a single mode is
supported in that direction. For given $D_y$ the wave number in $y$
direction is thus fixed to $k_y=\pi/D_y$ for \textit{all} modes.
Hence, for a given optical frequency, decreasing $D_y$ will reduce the
wave number $k$ in $z$ direction. At a certain threshold width, $k$
will vanish and the mode is no longer supported for widths smaller
than that threshold.

\begin{figure}[tb]
\includegraphics[width=18em]{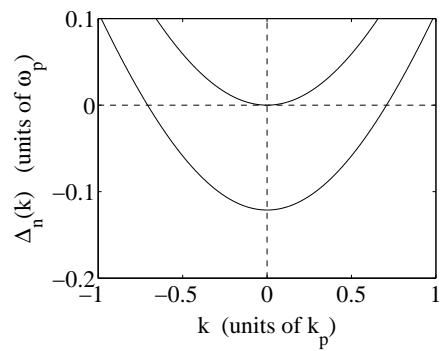}
\caption{Mode frequency $\Delta_n(k)$ versus longitudinal wave number $k$.
The parameters are $n_0=1.5$, $D_x=\lambda_p/\sqrt{n_0^2-1}$, and $D_y$ is
chosen such that the threshold of the first excited branch is $\omega_p$.}
\label{fig:modes}
\end{figure}

In Fig.~\ref{fig:modes} we show the numerically calculated frequencies
and longitudinal wave numbers of the modes supported by such a
structure. The width $D_y$ is chosen such that the threshold of the
first excited branch coincides with the pumping frequency $\omega_p$,
i.e., $\Delta_1(0)=0$. Hence, for all of the modes in this branch
$\Delta_1(k)\geq 0$ and a large atomic lightshift $L$,
Eq.~(\ref{eq:shift}), can be expected. Moreover, at threshold we have
$\frac{\partial\Delta_1(k)}{\partial k}=0$ and therefore a large
number of modes contributes nearly resonantly to the integral in $L$.
Thus, the presence of the microstructure largely enhances the mode
density near resonance.

An analytic approximation of Eq.~(\ref{eq:shift}) can be obtained
assuming that the transverse part of the mode functions is given by a unique
$f_n^{(T)}(x,y)$ within
each branch of modes for all the relevant frequencies, that is, assuming
constant values of $k_{x,n}$ and $k_{y,n}$ and therefore a constant transverse
wave number 
\be
q_n = \frac{1}{n_0}\sqrt{k_{x,n}^2+k_{y,n}^2}.
\ee
Introducing an exponential convergence factor to cut off high
frequencies which in the dipole approximation lead to an unphysical
logarithmic divergence, we find by complex contour integration
\be
L_n = -2\pi\frac{g^2 n_0}{c}|f_n^{(T)}(x_a,y_a)|^2
      \sqrt{1-\left(\frac{q_n}{r_n+is_n}\right)^2}.
\label{eq:shiftanal}
\ee
Here
\bea
r_n & = & \frac{1}{c}\left[-u_n+\sqrt{u_n^2+\kappa^2\omega_p^2}\right]^{1/2}, \\
s_n & = & \frac{\kappa\omega_p}{c}\frac{1}{r_n}, \\
u_n & = & \frac{1}{2}(c^2 q_n^2+\kappa^2-\omega_p^2).
\eea
For the parameters used in this Letter we compared these results with numerical
integrations of Eq.~(\ref{eq:shift}) and found excellent agreement.

\begin{figure}[tb]
\includegraphics[width=18em]{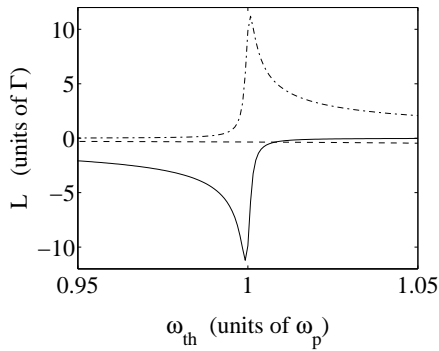}
\caption{Light shift $L$ versus threshold frequency $\omega_{th}$. 
The solid line is ${\rm Re}\{L_1\}$, dashed line
is ${\rm Re}\{L_0\}$, dash-dotted line is ${\rm Im}\{L_1\}$. The atomic
parameters correspond to the $D_2$ line of Rb, $\kappa = 0.001\omega_p$,
$n_0=1.5$, $D_x=\lambda_p/\sqrt{n_0^2-1}$. The atomic position is 
$(x,y,z)=(D_x/2,D_y/2,0)$, i.e., the point of maximum coupling at the surface
of the dielectric medium.
}
\label{fig:shift}
\end{figure}

An example for the light shift $L$ is depicted in
Fig.~\ref{fig:shift}. We plot the real part of the contribution $L_0$
from the lower energy branch of electromagnetic modes (the imaginary
part being approximately zero) as well as the real and imaginary parts
of $L_1$, the contribution from the excited states, versus the
threshold frequency $\omega_{th}$ of the excited branch. For the
chosen parameters, $\omega_{th}$ is a nearly linear function of the
width $D_y$, which varies approximately for 10\% within the plotted
range.

We note that $L_1$ is approximately constant and below one atomic
linewidth.  Hence the light shift induced by a branch of modes far
above threshold is in fact insignificant. The second (near-resonant)
branch of modes, on the other hand, yields a large light shift. For
$\omega_{th}<\omega_p$, travelling wave solutions exist in the excited
branch at the pump frequency, and the light shift is dominated by its
real part leading to increased spontaneous atomic decay by enhanced
emission of photons into the 1D microstructure. For
$\omega_{th}>\omega_p$, no travelling solutions exist, $L$ is
imaginary, and the main effect is a shift of the atomic frequency. At
threshold and assuming $\kappa\ll\omega_p$, Eq.~(\ref{eq:shiftanal})
can be approximated by \be L_n = 2\pi\frac{g^2 n_0}{c}|f_n(x_a)|^2
\frac{i-1}{2}\sqrt{\omega_p/\kappa}.  \ee Hence, the maximum possible
light shift is determined by the ratio of the photon loss rate from
the microstructure to the optical frequency.  Thus, for our specific
example of a structure the limiting factor will be the reflectivity of
the metallic coatings on two sides of the dielectric surface. For
example, a reflectivity of 99\% yields a decay rate $\kappa\approx
\omega_p/1000$ as used in Fig.~\ref{fig:shift}. Both the magnitude
and the scaling of the calculated light shift is very different from
the one observed between metallic plates \cite{Walther}.

\begin{figure}[tb]
\includegraphics[width=18em]{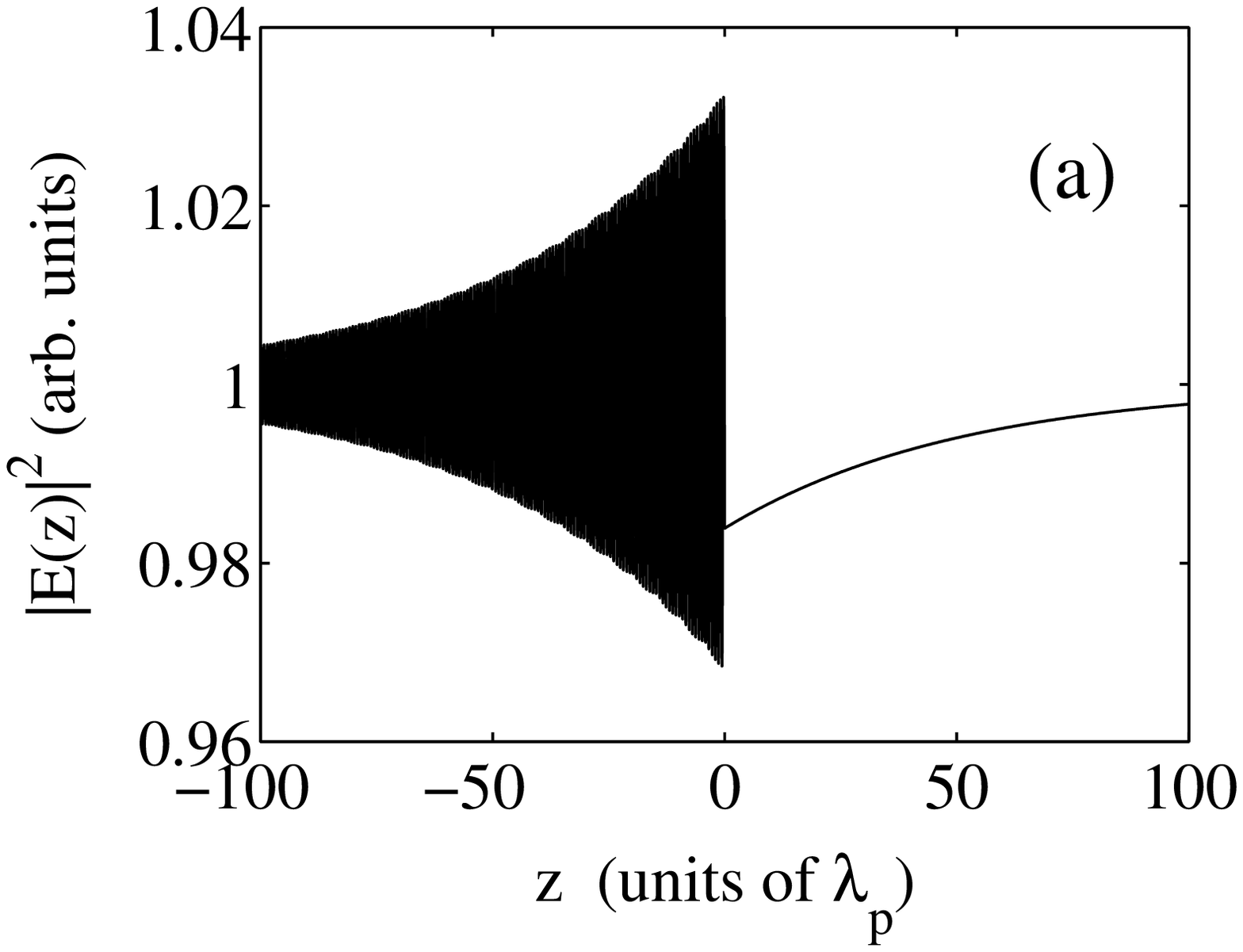}
\includegraphics[width=18em]{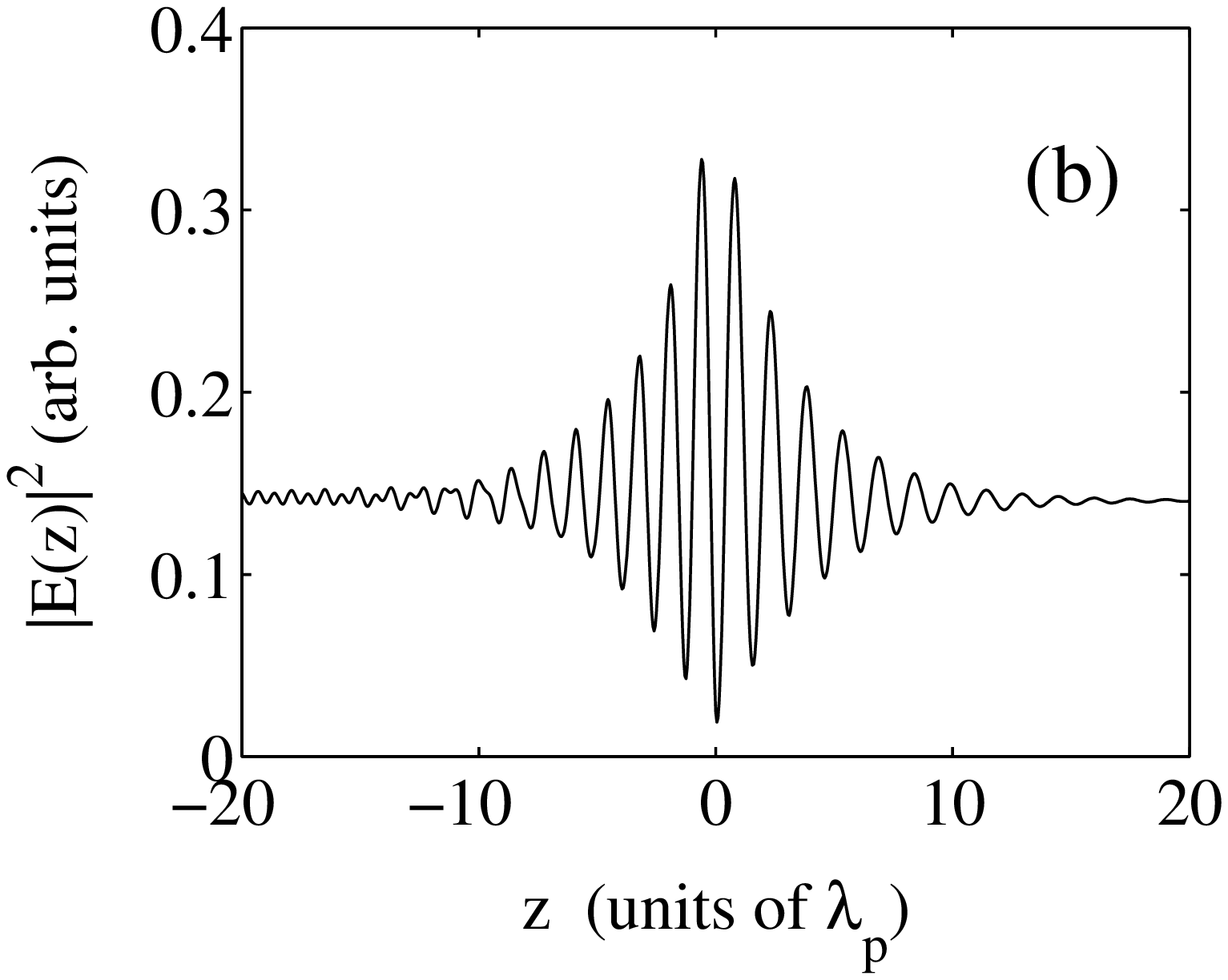}
\caption{Stationary light field intensity $|E(z)|^2$ along $z$ for $y=D_y/2$
and (a) $x=0$, (b) $x=D_x/2$. Parameters as in Fig.~\protect\ref{fig:shift}
with $\omega_{th}=\omega_p$, $\Delta_a=10\Gamma$.
}
\label{fig:field}
\end{figure}

Let us now address the question how this large atomic light shift could
be observed experimentally. Since the whole effect is due to the
strong coupling of the atom to the confined light modes, an obvious
possibility is to detect the backaction on the light field. Assuming
that only a single travelling wave with wave number $k_0$ of the lower
branch of modes is pumped and with the same simplifications as used to
obtain Eq.~(\ref{eq:shiftanal}) yields the following stationary
electric field (\ref{eq:field}):
\bea
E(\mathbf{x}) & \propto & e^{ik_0z}f_0^{(T)}(x,y) \nonumber \\
  & & -L_0\frac{e^{n_0(ir_0-s_0)|z|}}{i\Delta_a-\Gamma+L} f_0^{(T)}(x,y) 
     \nonumber \\
  & & -L_1 \frac{e^{n_0(ir_1-s_1)|z|}}{i\Delta_a-\Gamma+L} 
       \frac{f_0^{(T)}(x_a,y_a)}{f_1^{(T)}(x_a,y_a)}f_1^{(T)}(x,y),
\label{eq:fieldstat}
\eea 
where the first line gives the field of the single pumped mode,
the second line is the field of the light scattered into the lower
branch of modes, and the third line is the field scattered into the
upper branch. As an example we plot the field intensity along the $z$
direction at the center of the medium, Fig.~\ref{fig:field}(a), and on
the surface, Fig.~\ref{fig:field}(b).

At the center of the structure all modes of the excited branch vanish
and the electric field is formed by the lower branch only. Since the
light shift $L_0$ according to this branch is small, the change of the
electric field is small too. However, we see that the atom (at
position $z=0$) scatters some light from the pumped mode into its
degenerate counter-propagating mode. Hence, on top of the constant
intensity of the pumped mode, there appears a standing wave structure
on one side of the atom. Due to the damping of the light modes, this
standing wave has an exponentially decaying envelope with a decay
distance of $d_0 = c/(n_0\kappa)$. The electric field at the surface,
on the other hand, is dominated by the large light shift $L_1$ due to
the excited branch of modes and therefore has a much larger change of
amplitude, see Fig.~\ref{fig:field}(b). Similar enhancement of light scattering
has been predicted for a dielectric wire in a metallic wave guide
\cite{Saenz}.
According to Eq.~(\ref{eq:fieldstat}), we find
again an exponential decay with an approximate decay length of $d_1 =
c/(n_0\sqrt{\kappa\omega_p})$ which is much shorter than that of
Fig.~\ref{fig:field}(a). A spatially resolved detection of the photons
lost through the coatings of the dielectric structure would thus
reveal the significant change of the electric field intensity and
would serve as an implicit measurement of the enhanced atomic light
shift.

An alternative method to detect the light shift could come from an
experiment where a cloud of cold atoms is dropped onto the
microstructure and reflected by the evanescent light field. If the
atomic cloud is dilute enough such that the mean distance between the
atoms is larger than $d_1$, each atom will be scattered individually
and the reflection of the cloud will essentially be specular. On the
other hand if atoms are closer than $d_1$, they will interact with a
distorted light field as shown in Fig.~\ref{fig:field}(b). Since the
modulation of the light intensity along $x$ is roughly of the order of
the total intensity and since the periodicity is of the order of an
optical wavelength, the forces along $x$ will be comparable with the
force in $z$ direction. Hence the average reflection of a cloud of
atoms will be highly diffusive in this regime.

%%%%%%%%%%%%%%%%%%%%%%%%%%%%%%%%%%%%%%%%%%%%%%%%%%%%%%%%

\textit{Conclusions.- }
We have chosen here to discuss the interaction of an atom with a quasi
1D optical waveguide. However, our theory is easily applicable to
atom-photon interactions in a broad range of micro-optical structures.
Corresponding experiments can have impact on our basic knowledge about
atomic structure and quantum electrodynamics \cite{Hinds}. On the
practical side, the strongly enhanced atom-photon interaction could be
the basis of a new generation of micro-optical devices involving only
very few atoms and photons.

%%%%%%%%%%%%%%%%%%%%%%%%%%%%%%%%%%%%%%%%%%%%%%%%%%%%%%%%

%\acknowledgments

We thank J.\ Weiner, J.\ Schmiedmayer, and R.\ Folman for fruitful
discussions.  This work was supported by the Austrian Science
Foundation FWF (project P13435-TPH). P.~D.~ acknowledges the financial
support by the NSF of Hungary (OTKA F032341).

%%%%%%%%%%%%%%%%%%%%%%%%%%%%%%%%%%%%%%%%%%%%%%%%%%%%%%%%

%\vspace{5mm}

\end{document}